\documentclass[preprint,5p,twocolumn]{elsarticle}
\usepackage{graphicx}
\usepackage{xcolor}
\pdfoutput=1
\usepackage[numbers]{natbib}
\usepackage{color}

\usepackage{amssymb}
\usepackage{hyperref}
\usepackage{booktabs}
\journal{TBD}
\begin{document}
\title{Effective Model with Personalized Online Teaching and Learning Science in the Era of ChatGPT}
\address[1]{Department of Physics and Astronomy, College of Staten Island, Staten Island, New York 10314, USA}
\address[2]{Department of Physical Science, Bergen Community College, Paramus, New Jersey 07652, USA}
\author[1]{Kalani Hettiarachchilage}
\author[2]{Neel Haldolaarachchige}
\date{\today}

\begin{abstract}
The recent development of science education leads educators to explore new teaching and learning methodologies and restructure classes and assignments to bring students' knowledge to the highest level of education by allowing learners to gain various skills that are reflected in their future. 
We are discussing how to develop and deliver effective online courses by personalizing the class and individualizing the assignments for learners to succeed in Physics. Various teaching and learning activities are aimed at effective learning and interaction to enhance student learning in this class structure. The students' performances and progress are analyzed in four large introductory Physics classes. Student registration and retention are well preserved by providing learners freedom of education, a sense of belonging, mindset-growing encouragements, effective feedback, problem-solving strategies, and one-to-one communications. Academic integrity is well-balanced by individualizing the assignments, personalizing class materials, breaking down problems, and investigating ChatGPT usage. The students' performances are analyzed by evaluating class performance in total grade distribution, missing assignments, and choice-based learning. Student progress is investigated by analyzing their commitment, respect, willingness to improve, and time management toward education. Overall, the class model demonstrates great performance, progress, and achievements that indicate a positive learning environment for all students throughout the semester. Hence, the effective model with personalized assignments, and other strategies that we share will be highly beneficial for science education.
\end{abstract}


\maketitle
\section{Introduction}
Introductory-level science and mathematics classes are commonly offered through traditional in-person modality. Most of the students are coming from different fields of science, health care professionals, science education, business management, and engineering-related disciplines. Particularly introductory-level physics classes are the most challenging classes that students have to complete for their major requirements. On the other hand, these are the classes where learners can learn and develop many life-learning skills and strategies that will be reflected in their future. Therefore, class designing and restructuring of these key classes should be carefully handled by reflecting course learning and student learning outcomes. Various online modalities of these classes are implemented and tested in many colleges. Among them, online synchronous (virtual meetings as scheduled), online asynchronous (self-faced or scheduled), and hybrid (fused with in-person and virtual) are the key structures ~\cite{huangqing2021online, guo2020synchronous, faulconer2018comparison}. Following a well-structured semester schedule for all modalities is possible, but there are advantages and disadvantages for learners and instructors in all mechanisms in technical subjects like Physics. The main concerns of all types of online modalities are assessment methods, strategies, and academic integrity ~\cite{sutadji2021adaptation, hasan2021analysis, haldolaarachchige2022comparison}. Although there are various lockdown browsers~\cite{teclehaimanot2018ensuring, kuppers2017beyond} with enhanced features that can be used for online classes, still there are some difficulties in the assessment process.~\cite{garcia2021transformation, davis2019online} Especially, in Physics classes, students have to solve detailed problems followed by constructing diagrams, writing and deriving mathematical formulas, applying critical processes for algebraic and numeric solutions, scanning pages of work, and submitting to the learning management system. Among all modalities, synchronous class modality allows us to perform the assessment process intensively by solving some difficulties and mimicking real in-person learning experiences by enhancing diversity, equity, social justice, cultural response, engagement, attention, and interaction. 

In this paper, we discuss how to restructure a large-scale introductory physics course into an online effective deliverable by discussing the detailed development process and analyzing student performance. We study students' learning outcomes, performances, and progress throughout the process and achieve expected course learning outcomes and statistical representation of students' performances. We use academic dishonesty-minimizing techniques for all assessments, investigate ChatGPT-generated answers~\cite{chatgpt, yeadon2023death, kortemeyer2023could, bitzenbauer2023chatgpt, kuchemann2023physics, macisaac2023chatbots}, and provide detailed feedback to learners to learn from mistakes. Further, we provide learners with a task list to complete each part of the course and explore the improvement of students' commitment, interaction, knowledge, and skill development throughout the semester. We distribute surveys anonymously to check learners' feedback on their learning and experiences. 

Even though rapid improvements in online teaching and learning with the effect of COVID in recent times, there are many questions that remain particularly in online courses on introductory science. Now with the effect of ChatGPT, a lot of new questions are arising about online education. Our effort here is to address some of the selected and interesting research questions and our investigation and data made it possible to answer the following questions.

\begin{itemize}
 \item Is it possible to structure large online courses in an effective way for learners to learn and interact with the learning process?
 \item How to make personalized and sustainable assignments in large online courses?
 \item How to balance academic integrity with personalized online assessments?
 \item Can students effectively learn from online classes by developing critical thinking ad problem-solving skills while gaining conceptual knowledge? 
 \item How do personalize online classes affect student registration and retention?
 \item How do the students' demographics relate to the student's performance on online courses? 
  \end{itemize}
  
\section{Methods}
There are four different courses that we restructure and analyze. All four courses are introductory-level physics courses at the College of Staten Island (CSI), within the system of the City University of New York (CUNY), United States of America. CSI is a mid-size 4-year university with an enrolment of about 11000 undergraduate students. These 4 introductory physics courses are generally considered service courses, designed for non-physics major students and offered by the Department of Physics and Astronomy. 

\begin{itemize}
    \item Physics I (PHY116) class is an undergraduate algebra-level physics course similar to the college Physics I course in any college/university in the United States. This is a required course for students majoring in biology and  health sciences. 
    \item Physics II (PHY156) is the second sequence of a two-semester undergraduate algebra-level physics course required for most students majoring in health sciences. 
    \item Introduction to Physics (PHY114) is a conceptual physics course required for students majoring in life and physical sciences, especially for the nursing program. 
    \item Nature of physical processes (PHY206) is also a conceptual physics course required for students majoring in some liberal art subjects and education science majors.
\end{itemize}

\subsection{Class Structure}
We restructure the class into three components to provide learners with the best and equal education and learning experience as discussed below. 
\begin{enumerate}
    \item Asynchronous Lecture: The goal of this part of the course is to provide the key content knowledge of the crucial concepts that will use during the rest of the class structure. E-lecturing is assigned to 55 minutes period of the class. Learners are supposed to complete a given task while watching the recorded lectures and submit it before they come to the recitation problem-solving session. E-lecture gives freedom to learn the material with an individual schedule. Learners can expose themselves to the subject matter by exploring crucial concepts that needed to apply for problem-solving. E-lecturing allows them to revisit and review class material as much as needed.
     \item Synchronous problem-solving: This component is assigned 1 hour and 55 minutes of the class period to use the learned concepts in E-lectures toward problem-solving and real-world applications. This component of the course is very important for students to learn how to apply what they learn and develop problem-solving strategies by working with a given set of problems. At the end of the class, completed work is supposed to be submitted for a completion grade. The technique was investigated under mandatory attendance policy and choice-based attendance policy in two separate semesters to check the progress and effectiveness. The problem-solving recitation session allows learners to build a bridge between what they heard and learned in E-lecture to real-life problem-solving and applications by making sense of the content and its relevance.
    \item In-person laboratory:  This is where we bring students in person and  connect physics concepts to laboratory experiments to provide them with hands-on experiences. The lab is used to introduce, reinforce, and/or enrich the treatment of related topics studied in the lecture portion. Learners can visualize, demonstrate, and analyze physics concepts while developing hands-on experiences, collaborations, communications, and organization skills. Learners should participate in a registered lab session and conduct laboratory exercises and submit required work as assigned by the instructor. Students must attend and complete 80\% or more of the required lab work to pass the course.
\end{enumerate}

\subsection{Learning Management System (LMS)}
\begin{figure}[!htbp]
  \centerline{\includegraphics[width=0.5\textwidth]{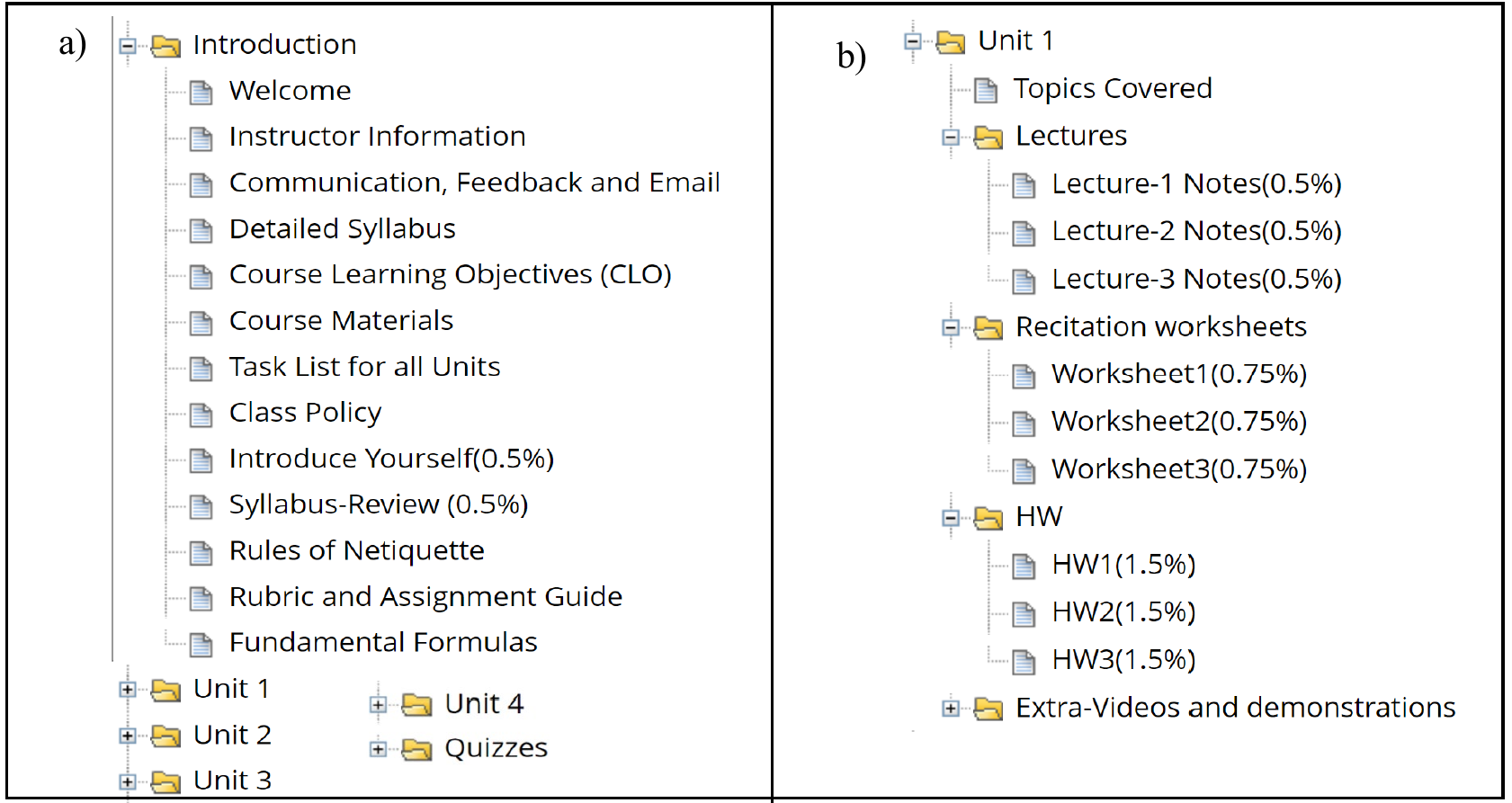}}
  \caption
    {(Color online) Structure of Blackboard class page. a) a General view of the Blackboard navigation with orientation tab and unit module with well-structured content for learners to engage in online classes, and b) a detailed view of the unit module.   
   }
\label{BlackBoard}
\end{figure}

We use the Blackboard learning management system effectively to engage learners in various and essential activities of the course ~\cite{Blackboard}. This is structured in a way that learners can elevate their education through a series of exercises, individually and in groups that are started with easy access to conceptual video lectures, guided problem-solving video lessons, assignments, surveys, detailed grade book, running grades, instructor's feedback, discussion forums, Padlets~\cite{deni2018padlet, rashid2019using}, voice threads~\cite{salas2015value}, and many other resources. The blackboard page is mainly structured into six tabs as in Figure~\ref{BlackBoard}. The introduction tab will orientate students to the class with a warm welcoming video massage and class overview: blackboard navigation, class structure, office hours, meeting times, course learning goals, communication methods, instructor information, a detailed syllabus, a submission task list, introduction forum with Padlet, a syllabus review quiz, rules of Netiquette and class policies. The class materials are distributed to four separate tabs as unit modules. Each unit includes topics covered, an E-lecturing folder with three lectures and related assignments, a problem-solving folder with detailed problem-solving videos and three worksheets to complete, a homework folder with all three homework assignments, and an optional extra folder with other supported materials. Additionally, anonymous surveys are posted from time to time. All quizzes and exams are posted under the quiz tab outside the four Unit folders for students to easily navigate. 

\subsection{Education Resources}
The courses are based on an open educational resource (OER) textbook ~\cite{urone2012college}. The plan is to make the course more affordable with no extra cost for learners. The OER textbook provides guidance to learners to explore knowledge of materials, however, this OER textbook is for algebra-based physics and not well suited for conceptual physics therefore we had to do certain changes and provide specific instruction with footnotes for most of the chapters. E-lectures are designed closely following the OER textbook but are mostly self-designed by the instructor.~\cite{Kalani} 
Other course materials (worksheets, homework, quiz/exam questions) are carefully designed from scratch by the instructor. All assignments are updated every semester to reduce academic dishonesty and be sustainable for future semesters.
All assignments are graded digitally by hand through Blackboard without using expensive auto-graded homework systems.

\subsection{Grading scheme} 
The grading items are shown in table~\ref{tab:Grade}. For each unit, learners have to submit 3 lecture notes, 3 worksheets, 3 homework, and a quiz. All together 12 lecture notes, 12 worksheets, 12 homework, 12 lab work, 4 quizzes, and a cumulative final are assigned for the semester. We combine the lecture notes assignment with homework in Fall 22 but separated them in Spring 23 to enhance the learning from E-lecturing. We follow the same grading criteria for all items in both semesters. Lecture notes and worksheets are ungraded assignments (which means students get full credits for submissions by completing and following the given format and the model). All homework and quizzes are graded by following the rubric in Figure~\ref{Rubric}. All lecture assignments excluding lab work should be handwritten by students and graded digitally through the blackboard by following the given rubric by the instructor.

\begin{table}[!htp]
\caption{\label{tab:Grade} Grading items for final class grade calculation}
\begin{tabular*}{\columnwidth}{@{\extracolsep{\fill}}lc}
\toprule
Item description& Score(\%)\\ 
\hline
4 unit Quizzes: 8 points each& 32 \\ 
Final Exam (cumulative)& 10\\
12 Homework assignments: 1.5 points each& 18\\
12 Worksheet completion: 0.75 points each$\:$& 9\\
12 Lecture note: 0.5 points each$\:$ &6\\
Syllabus review quiz & 0.5\\
Discussion forum and Padlet & 0.5\\
12 Lab work: 80\% completion to pass the class& 25\\
\hline
Total& 100\\
\hline
\end{tabular*}
\end{table}

\subsection{Type of Assignments}
\subsubsection{Lecture note assignment}
\begin{figure}[!htbp]
  \centerline{\includegraphics[width=0.5\textwidth]{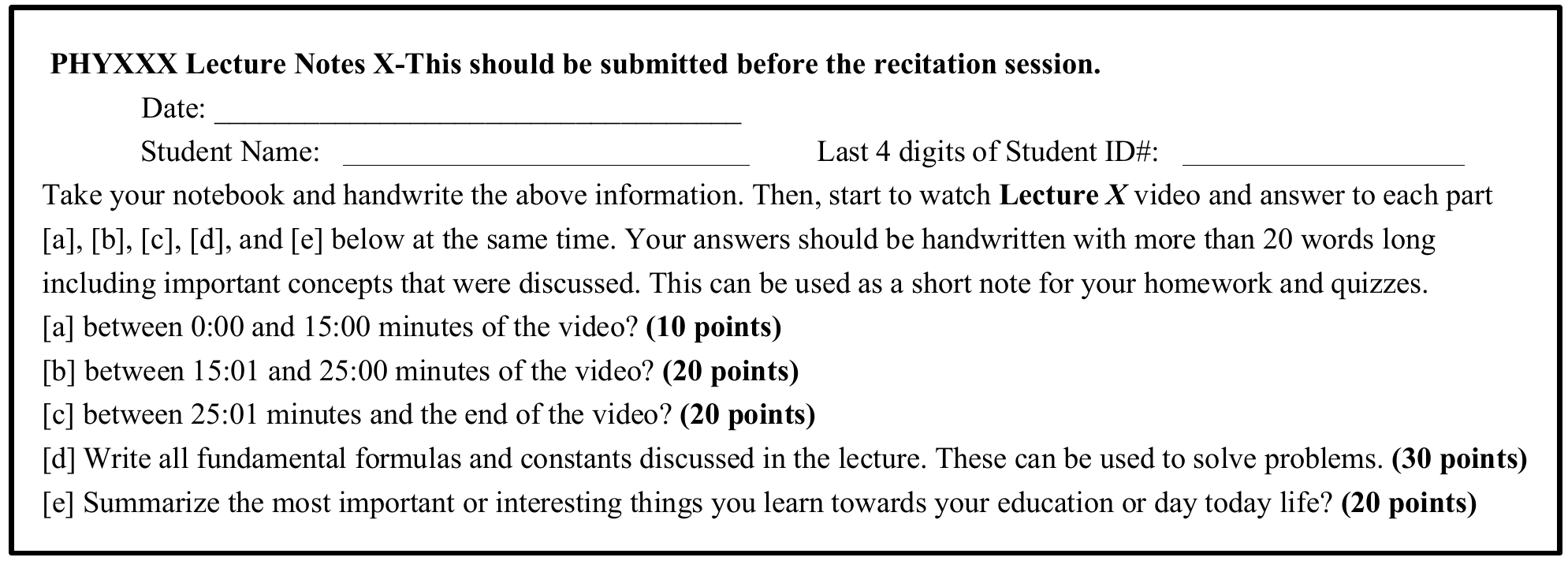}}
  \caption
    {(Color online) Lecture notes task. Learners suppose to complete this simple task in their notebooks by watching the E-lectures. 
   }
\label{LectureNotes}
\end{figure}

Each E-lecture describes the subject matter with crucial concepts and derives or introduces mathematical formulas that will be used for problem-solving. 55 minutes of class time is assigned for each lecture although the E-lecture is shorter than that. The learners get an opportunity to revisit the lecture or refer to the textbook for more information on materials. A small task for each lecture as shown in Figure~\ref{LectureNotes} is given as lecture notes. This assignment has to be completed while watching the E-lectures. Students have to hand-write the answers in their notebooks by following the model and submit it to Blackboard before the problem-solving session. 

The idea of this task is to expose students to learning concepts, physical principles, and formulas through E-lectures. Students have to make their own short lecture notes. Next, we encourage them to use their own notes during the problem-solving session, homework, and quizzes. This is how we personalize the class by providing students with a very helpful way to learn concepts, principles and laws, formulas, constants, units, and variables easily without using random picks from various resources. Since they will receive points as in table~\ref{tab:Grade}, all learners complete the task to their best. Learners will be able to learn concepts more deeply with this method rather than sitting in-person class with mostly passive learning. By restructuring the E-lecturing method, we create a bridge to overcome some difficulties by providing them with enough time to learn from an equivalent learning experience. In Fall 22 we asked learners to write lecture notes as the first question in the Homework. In Spring 23, we separate lecture notes from Homework, which will allow learners to study lecture notes carefully and more effectively before joining the problem-solving session.

\begin{figure}[!htbp]
  \centerline{\includegraphics[width=0.5\textwidth]{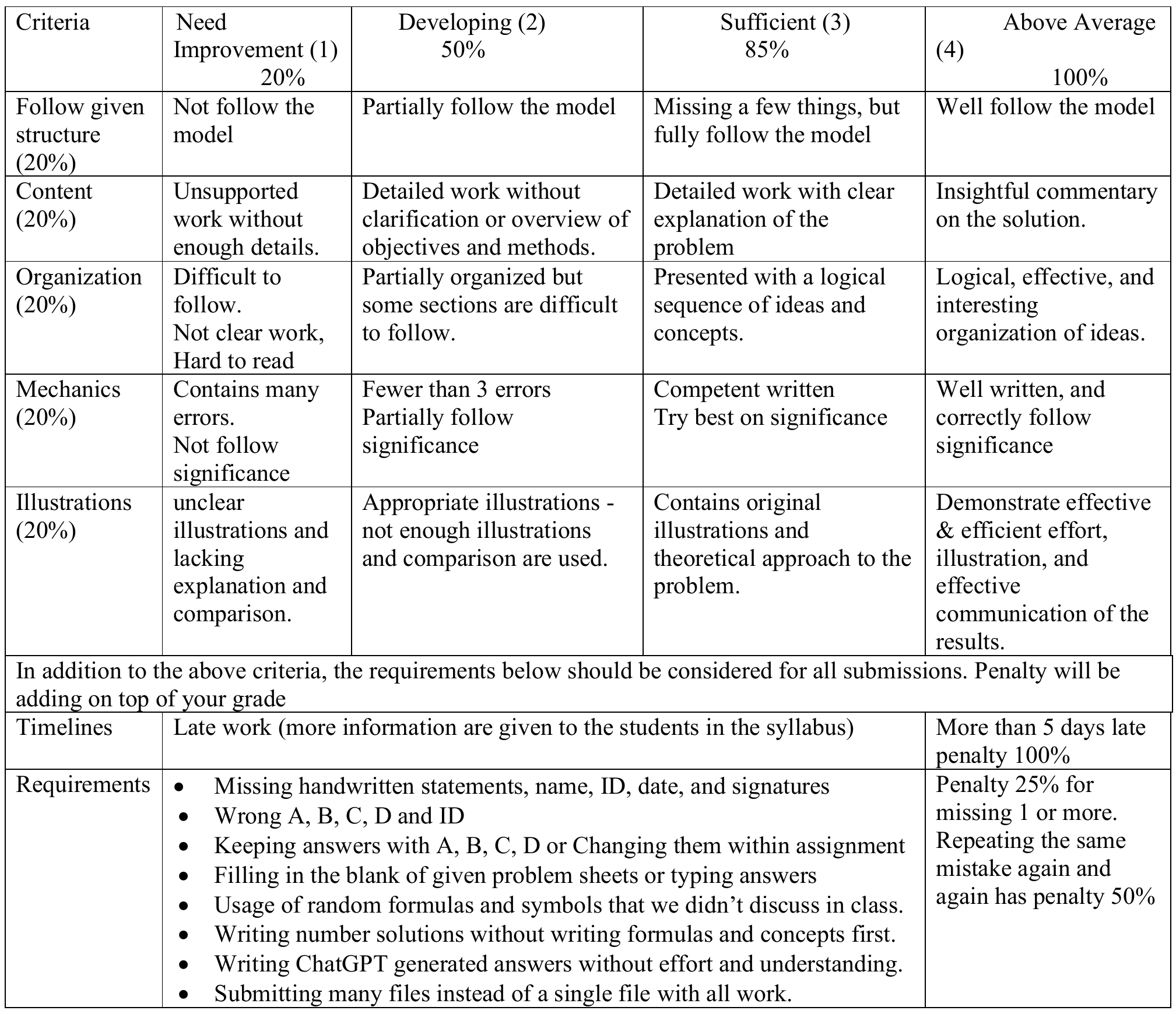}}
  \caption
    {(Color online) Rubric for grading homework, quizzes, and final exam. The rubric shows roughly the way that the grade will be assigned. The way grades are assigned depends on how well learners meet the goals of each assignment. Learners suppose to use the rubric as a guideline for preparing the assignments. 
   }
\label{Rubric}
\end{figure}

\subsubsection{Worksheet assignment}
After the successful completion of the E-lecture portion, we bring learners together for a problem-solving session (1 hour and 55 minutes). This is an enhanced version (not exactly the same) of a recent trend of virtual or web-based recitation models in most colleges/universities in the United States.~\cite{sundari2021interactive, diyana2020effectiveness, adila2019recitation} There is a worksheet with a problem set that helps learners to apply the knowledge gained from the E-lecture and learn how the formulas, symbols, constants, laws, variables, and concepts are applied to solve problems. This is another opportunity for them to think and to make sense of all the concepts learned in E-lecture. This session will provide learners to develop problem-solving strategies, critical and analytical thinking, and applications of what they learn. This is a discussion session, we walk through all problems with student engagement in break-out rooms or as a whole group. We work through all the important steps of problem-solving by starting with diagrams, labeling, fundamental formulas, units, significance, ...etc. After the session is completed, learners have to submit their work to the blackboard as an ungraded assignment to receive a completion grade as in table~\ref{tab:Grade}. This will help learners to complete their homework individually. We tested this session as mandatory attendance and also as a choice-based attendance and compared it with students' progress and performances (details in a later section). The choice-based version is very helpful for students who have concerns about the materials or difficulty with learning especially in this kind of technical subjects so that they can watch the recording by spending time on it. Further, the choice-based session is very helpful for students who work full-time or part-time since it provides flexibility in their schedules.

\subsubsection{Homework assignment with ID-integrated questions}
After completion of the worksheet, students have to complete the homework assignment for each lecture individually. They will get about a week to submit after the worksheet discussion, and homework questions are very similar to worksheet questions. Since we are not using expensive publisher homework systems, we have to grade homework submissions digitally through Blackboard. The grading is done by separating homework into two parts such as graded and ungraded sections. The graded part is worth 75\% and very detailed feedback is given for students to learn from mistakes. The ungraded part is worth 25\% for successful completion. All grading is done according to the rubric given in Figure~\ref{Rubric}.

\begin{figure}[!htbp]
  \centerline{\includegraphics[width=0.5\textwidth]{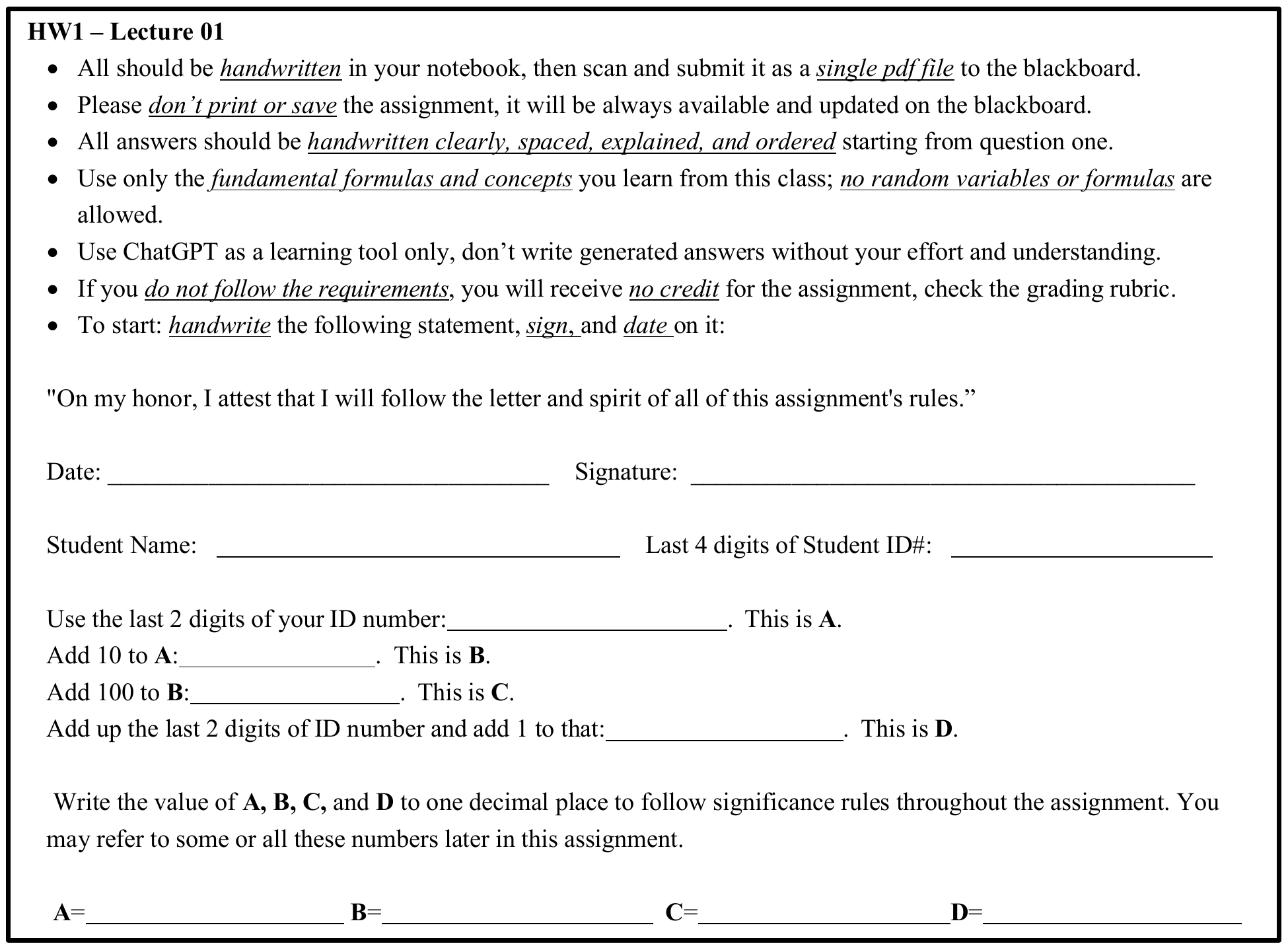}}
  \caption
    {(Color online) Student ID integration to the problems. All 12 homework, 4 quizzes, and the final exam are created by integrating the last 4 digits of the student's ID. Learners have to use \textbf{A, B, C,} and \textbf{D} in the calculation and conclude accordingly. This will create randomization and help students to work on their own problems and conclude differently.   
   }
\label{AssignmentID}
\end{figure}

Homework assignments are implemented with several different methods to minimize academic dishonesty issues. Homework assignments are integrated with Students' last four digits of the ID as shown in Figure~\ref{AssignmentID}. The ID-integrated questions will provide personalized questions for each student. This makes it not possible to share homework answers between students because every student gets a different answer. Therefore, this method forces students to work individually.  Also at the top of the paper student have to write the honor statement by hand as shown in Figure~\ref{AssignmentID}. This provides a psychological effect on students' minds to think that they have to do it individually, follow assignments requirements, and feel about their own work and learning. Video recordings of the homework requirements and the way of completing the assignments are posted on the blackboard which shows a step-by-step process. Further, all requirements are noted on the top of each assignment as in Figure~\ref{AssignmentID} and also restate with penalties on Figure~\ref{Rubric} to force students to follow the rules and learn better through the designed process of personalized learning. 

\subsubsection{Unit Quizzes with ID-integrated questions}
Unit quizzes are assigned to do at the end of each unit. The problems of the unit quizzes are similar to the worksheet and homework problems and are also integrated with the last 4 digits of the student's ID to vary the problem and individualize the conclusion. Quizzes are done synchronously and proctored via a video conference system for Physics I (PHY116) and Physics II (PHY156) classes.~\cite{haldolaarachchige2022comparison} Online asynchronous timed quizzes are given to conceptual physics classes (PHY206 and PHY114). Students' progress and learning are evaluated. Quiz questions are written by integrating symbols and breaking them down into parts to evaluate students' proficiency, understanding, learning of the materials, and problem-solving skills in each concept. The requirements are shown in Figure~\ref{AssignmentID}. Further, the student work is compared to ChatGPT-generated answers and checked with other students' answers. 

\subsubsection{Final exam with ID-integrated questions}
The final cumulative exam is designed to test overall knowledge by asking very simple and critical problems which will be the simple versions of quizzes, homework, and worksheets. The final exam is formatted with the same requirements and ID integration as shown in Figure~\ref{AssignmentID}. Online timed synchronous proctored finals are given to two-semester algebra-based physics classes:  PHY116 and PHY156. Timed asynchronous finals are given to conceptual physics classes: PHY114 and PHY206. 

\begin{figure}[!h]
  \centerline{\includegraphics[width=0.5\textwidth]{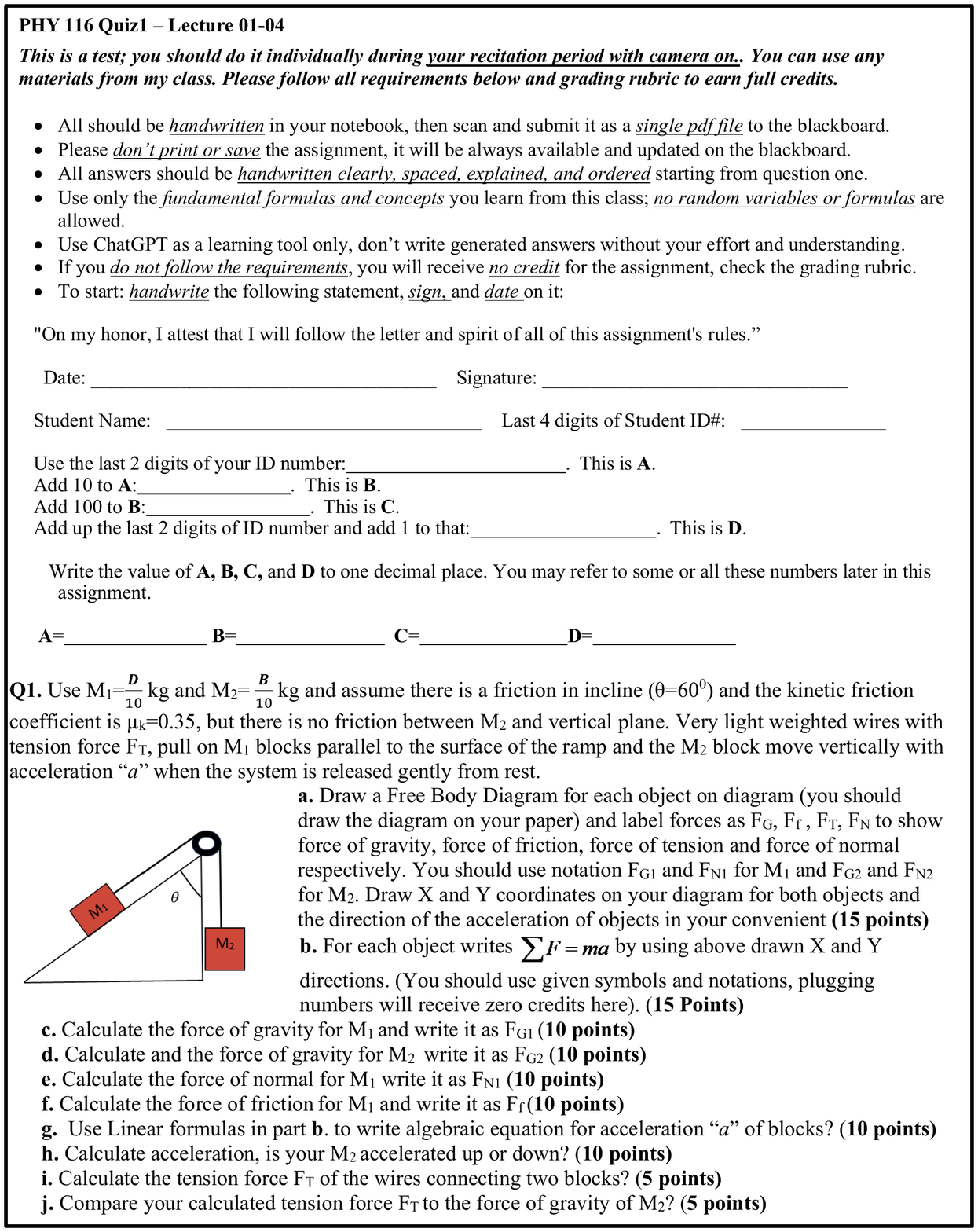}}
  \caption
    {(Color online) An example assignment of physics I (PHY116) class about Newton's laws of motion. This problem includes all assignment requirements and imposing concepts, critical thinking, and logical skills, personalizing to the class material by breaking questions into parts and cross-questioning about the conclusion and real application. 
   }
\label{Assignment}
\end{figure}

\begin{figure}[!ht]
  \centerline{\includegraphics[width=0.5\textwidth]{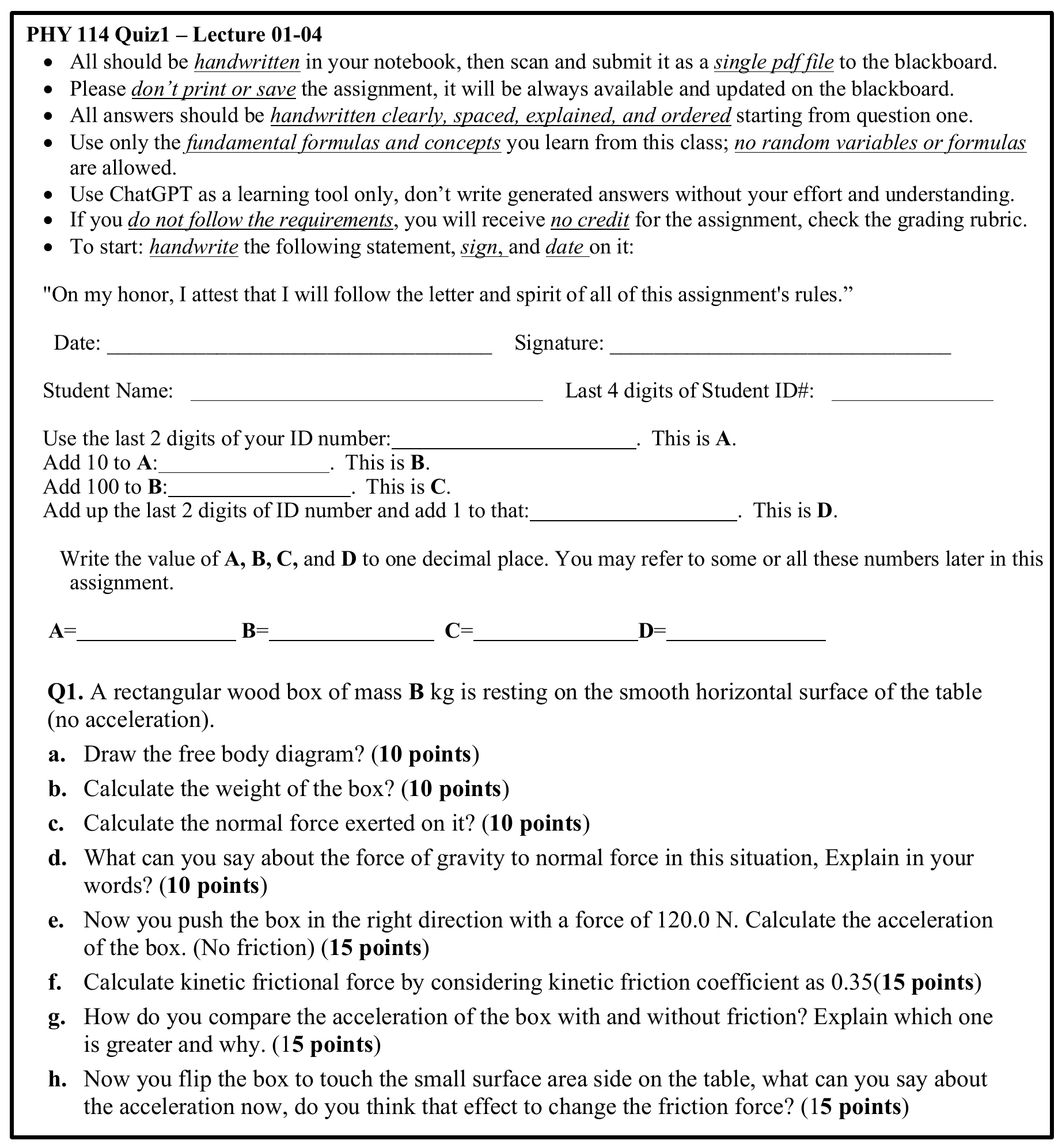}}
  \caption
    {(Color online) An example assignment for conceptual physics class (PHY114) for nursing and education majors about Newton's laws of motion. This problem includes all assignment requirements and focuses on concepts and measures their understanding on a conceptual level and logical applications.
   }
\label{AssignmentConcept}
\end{figure}

\section{Results and Discussion}
\subsection{Academic Integrity}
Student commitment to academic honesty, trust, fairness, respect, responsibility, and engagement is investigated throughout the semester in all assignments as shown in Figure~\ref{AssignmentID}. Asking to write the handwritten honors statement, name, ID, and signature on each assignment will force students to be honest with their own work. Accepting only handwritten answers make students rewrite and rethink what they learn, what they are writing, and what is wrong with their process. ID-integrated problems will personalize the problem to each student with a unique answer, making them think about the physical meaning of answers, and forcing them to personalize conclusions according to their own calculations. Breaking down problems into a few sections including conceptual, algebraic, applicable, logical, and critical objectives is allowed students to venture into their learning individually. Allowing students to use their own notes during synchronous proctored timed quizzes force learners to spend their time effectively on the given problems than trying to search online. Imposing assignment requirements and providing grading rubrics to follow the instructor model or expectations will help learners to personalize their work to the class without random symbols and formulas dig from online or ChatGPT.~\cite{macisaac2023chatbots, west2023ai, rudolph2023chatgpt} As instructors, we generate the ChatGPT answers a few times to each assignment even before posting the assignments. That will help us to restructure the questions to measure student learning other than digging for wrong or misleading answers provided by ChatGPT. Further, learners' work is graded by investigating students' effort and knowledge about the materials by comparing ChaGPT-generated answers. 

Figure~\ref{Assignment} shows the sample problem assigned to the learners by implementing the above-discussed remarks for the Physics I (PHY116) class. ChatGPT-generated answers for this question are misleading the students' process of answering such as a free body diagram, wrong X and y components, wrong directions or symbols for acceleration, not showing Newton's second law as discussed in the class, the difficulty of handling ID, misleading conclusions ...etc.  Figure~\ref{AssignmentConcept} shows the sample problem assigned to the conceptual physics class by implementing the above-discussed remarks. ChatGPT's answer to this question is more general and did not break down the answer logically to each part correctly and does not explain as the instructor personalizes teaching techniques. AI-generated solutions are very easy to identify. Also, by studying answers generated by ChatGPT we found that it is possible to do certain adjustments that ChatGPT will not be able to clarify because at present it solves the problems by using basic common knowledge extracted from the internet. For example, in the problem in Figure~\ref{Assignment}, ChatGPT gives the wrong x and Y components of the forces since the incline angle gives in a different way. ChatGPT always considers the given angle with respect to the +X axis and provides the general force components as $F_{x}$=F cos$\theta$ and $F_{y}$=F sin$\theta$ by taking the angle with respect to the +X axis. After careful investigation over two semesters, we found that ChatGPT is not a problem at present but it can be used as a learning tool. Additionally, since students have to follow requirements and ID integration for assignments, students really have to work and learn individually to succeed in the class. Students are informed about the penalties for the use of AI-generated answers for all assignments as Figure~\ref{AssignmentID} and Figure~\ref{Rubric}.

Also, we investigated the possibility of getting the answers for E-lecture assignments by using ChatGPT.\cite{chatgpt} We ask the question from ChatGPT exactly as it is given in Figure~\ref{LectureNotes}. Since it generates the responses based on its knowledge the response was that it is not able to answer because the information about the video does not exist in its knowledge. However, we found that if one copy-pastes the transcript of the video into the ChatGpt then it can give exact answers. Which means there is a challenge to academic integrity.~\cite{dawson2020defending, eaton2021plagiarism, holden2021academic} To minimize the academic dishonesty issue with this particular assignment we investigated students' responses against the ChatGPT-generated responses to the video transcript. After careful investigation, we found that there were no issues of academic integrity. It is important to mention that every assignment consists of a range of penalties for academic dishonesty and particularly if a student's answer matched with AI-generated responses. We think that it is very important to inform students about instructors' knowledge of any new developments such as ChatGPT. This can make psychological effects on students' minds against the possibility of academic dishonesty. 

\subsection{Learners demographics and diversity}
\begin{figure}[!h]
  \centerline{\includegraphics[width=0.5\textwidth]{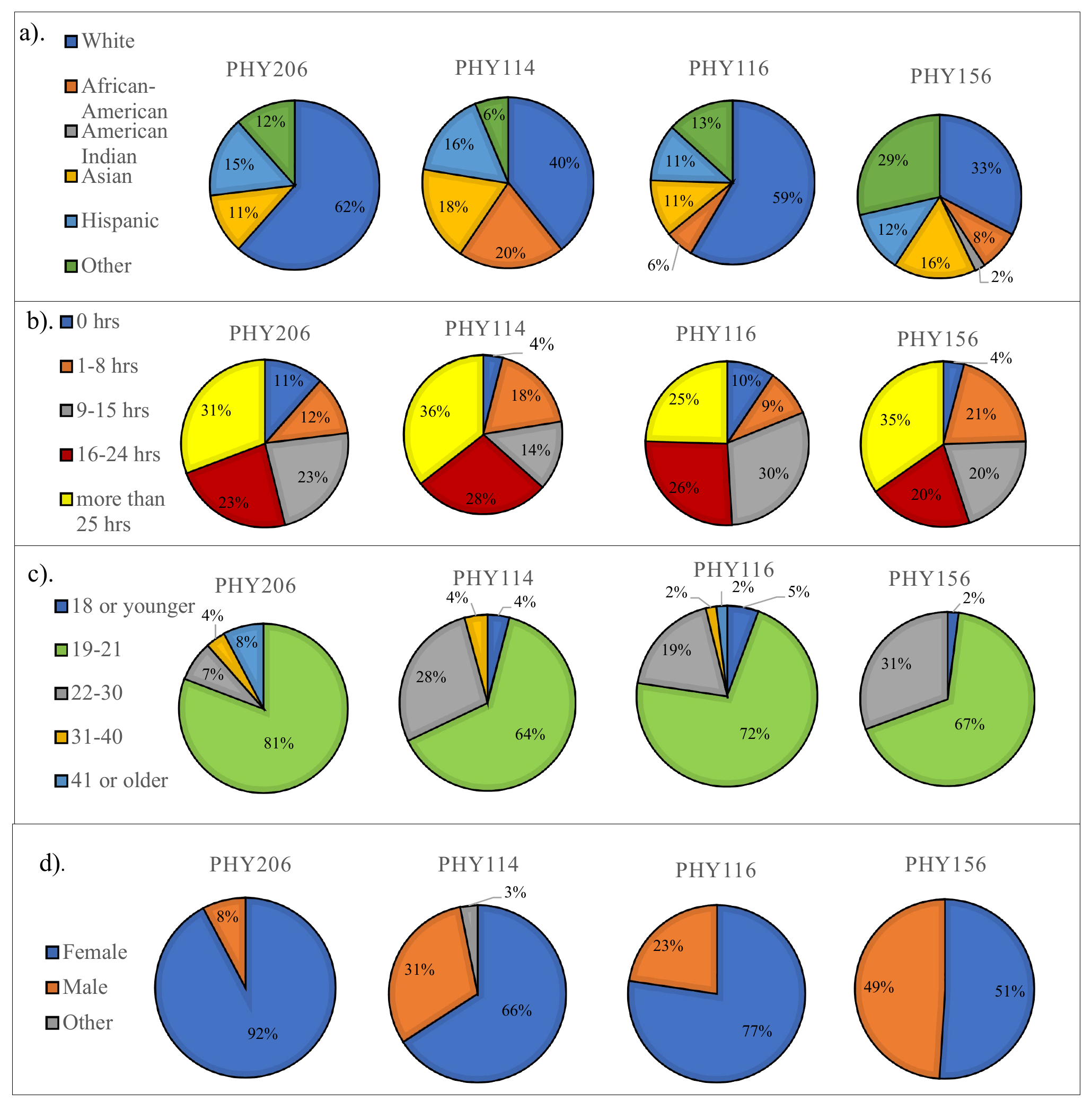}}
  \caption
    {(Color online) Learners' demographics and diversity for four different undergraduate physics classes. Panel a) displays the ethnic group, panel b) indicates the working load other than studies, panel c) displays the age group, and panel d) represents the gender group of the learners}
\label{Demog}
\end{figure}

All data are collected from the College of Staten Island (CSI), a senior college (or 4-year university) within The City University of New York (CUNY). The college has a very diverse community in terms of  demographic, age, gender, working and learning together, and language difficulties. Most learners are from the New York City area with less than 1\% of out-of-state and less than 1\% of out-of-country.~\cite{CSI}. Data collected from the four different physics classes are shown in Figure~\ref{Demog}. Data is collected total of about 800 registered students:  250 students in PHY116, 200 students in PHY156, 250 students in PHY114, and 100 students in PHY206. More than 75\% of them volunteered to complete the anonymous surveys provided by the Blackboard learning management system. Learners' ethnic group is dominated by white Americans in all four classes and the African-American ethnic group is yield in the educational major program (see in Figure~\ref{Demog} panel a)). Working and learning together students are dominated by all four classes, around 30\% or more will be working more than 25 hours per week other than studies. Most learners belong to the 19-21 years age group and are dominated by females in all four classes (see in Figure~\ref{Demog} panel c) and d)). Learners' demographic study suggests that this new model (3 component class structure) is more suitable for the student population at CSI because a large percentage of students in all 4 different introductory physics classes are working (full or part-time) while taking a full-time course load.~\cite{noack2009student, salehi2019demographic, van2020equity}

\subsection{Student registration and retention}
Student registration for the above four undergraduate physics classes is usually in high demand because the classes are required to complete for different majors. The classes are usually filled with health science professionals, nursing, and educational science majors. Registration has been elevated a lot with this online class structure since learners have the freedom to take at least part of the class online at their own pace, especially most students working full-time as shown in Figure~\ref{Demog} panel b). The number of students who enroll mostly stayed in the class and completed it successfully. Student retention average is 99\% for PHY156 and PHY206, 90\% for PHY116, and 85\% for PHY114. These averages are well above the normal retention levels for the in-person class model. The variation between different classes can be easily understood by looking at students' requirements and experiences of taking the class. PHY156 is a required second-semester physics course for those who completed PHY116 earlier. These learners have already experience in Physics I and the class structure, they are ready to work hard and learn to complete the course. PHY206 students will stay to complete the class since it is a required course. These students are future educators who know the value of education and instructor effort. Further, these students are mostly focused on education, and around 11\% are not engaging in full-time jobs as in Figure~\ref{Demog} panel b). PHY116 and PHY114 students register for the class by knowing they need to complete the course as a requirement, but some of them are unable to stay focused and complete assigned work on time due to external and conflicting work. Also, this may be the very first and most challenging science course that they are taking. Therefore, some percentage may drop out in the middle of the semester and usually re-register in the future semester. Retention is one of the main concerns, particularly from the administration's point of view. This new course structure shows how to keep very high retention rates while keeping a high standard of teaching and learning.~\cite{akanbi2021effect, wingate2018impact, timonen2022learning}

\subsection{Assessments and students' performance}
This part of the study shows the analysis of student performance on the current model but we are unable to compare students' outcomes of this class structure to in-person structure due to the following reasons, a) this model started just after the covid pandemic (2020-2021) and at the present CSI does not offer any of these classes in-person, b) before covid in-person classes were taught by different instructors with 18-36 students in each class, c) within in-person classes the instructors had the freedom of teaching materials and evaluate the students according to their own criteria, d) within in-person classes they have the freedom to use various resources (textbooks, homework systems, and online test generation tools). However, the restructured fully online present class model is totally new, all students are assigned to one single online lecture portion as a large class (100-150 students in each), and the lecture class was done by a single instructor with zero-cost (OER) materials by providing equal opportunities and educational experiences for all students. Therefore, we have to omit the comparison of performances or progress between the current and in-person models. 

\begin{figure}[!h]
  \centerline{\includegraphics[width=0.5\textwidth]{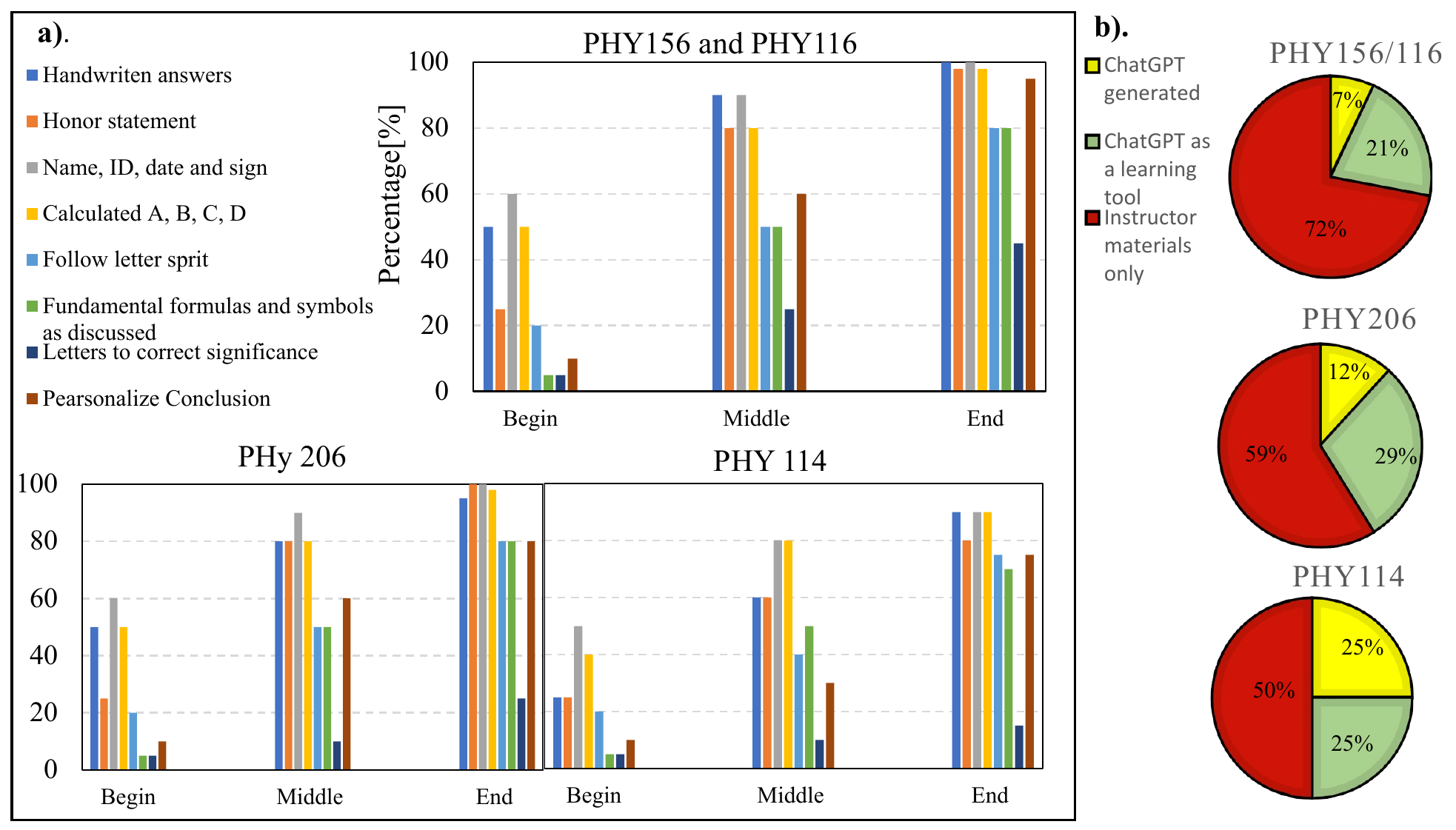}}
  \caption
    {(Color online) Learners' outcomes of following instructor personalize guidelines and requirements for the four different classes. Panel a) displays the progress of the learner's effort in following requirements with instructor feedback throughout the semester. Panel b) indicates the percentage of ChatGPT-generated answers, usage of ChatGPT as a learning tool, and reliance upon only class materials.}
\label{FollowGuide}
\end{figure}

First, we investigate the students' efforts of following instructor guidelines, models, and requirements for all assignments throughout the semester. Assignment requirements shown in Figure~\ref{AssignmentID} and grading rubric shown in Figure~\ref{Rubric} are provided to the learners as discussed above. Further, they were reminded from time to time through the LMS announcements and digital feedback with graded work. A recorded video with all details about assignments and the completion process is provided at the beginning of the semester as an "Assignment Guide". With about 800 students in four different classes and two-to-three semesters of data, we investigated the student respect of the above-mentioned objectives. Since there is no variation in all semesters, all objectives are concluded together in Figure~\ref{FollowGuide}. Panel a) of Figure~\ref{FollowGuide} represents the semester progress of the objectives for all four classes: PHY156 and PHY 116 represent together on the top, PHY206 indicates the bottom left, and PHY114 shows the bottom right. Writing handwritten answers started from around 50\% for PHY116, PHY156, and PHY206 and improved to 98\% for both PHY116 and PHY156 and 80\% for PHY206. PHY114 started from around 25\% and improved to 85\%. Writing honors statements started from around 25\% for all four classes, and improved to 98\% for all classes except PHY114 reached below 80\%. Writing the name, ID, and signature started from around 60\% for all classes, and improved above 95\%. Writing ID number integrated letters (A, B, C, D) started from around 50\% for all classes and reached above 90\%. Following ID integrated letters for all classes started very low about 20\% and improved to 80\% except PHY114 only reached 65\%. Use of given fundamental formulas and symbols started from less than 10\% for all classes and improved to 80\% for PHY116 and 156 but around 60\% for PHY206 and PHY114. Writing answers with correct significant figures started very low at 5\% and didn't observe much improvement from PHY206 and PHY114, but better improvement was observed in PHY 156 and 116, but still needed to be improved in future semesters. Writing personalized conclusions according to students' own calculated answers started at around 8\% and improved to above 80\% for PHY 116 and 156 and above 60\% for the other two classes. The results suggest that the following requirements depend on the class and mostly improved during the semester with effective feedback on grading. All requirements are preferably followed by PHY116 and PHY156 and reluctantly followed by PHY114. The range of improvements seems to have a correlation with the learner's discipline. 

In Figure~\ref{FollowGuide} panel b), we test the learners' usage of the instructor's own materials for their assignments compared to chatGPT usage as cheating and learning. In the figure ChatGPT-generated means that some students used ChatGPT to cheat (just to find the answers for homework and other assignments). In the figure ChatGPT as a learning tool means, learners actually check the answers on ChatGPT, but really personalize to the class with given instruction. For the second category of students, we can imply that they use it to learn and test their approach. In the figure, instructor materials only mean learners are following only the instructor materials and guidance and assume no usage of ChatGPT. The studies are completed by checking a few important points observed during the grading such as steps, processes, wordings, definitions, and formulas in written answers.~\cite{khan2021student, tunggyshbay2023flipped, gunawan2023online, rahayu2022effectiveness, agustin2023enhancement}
According to Figure~\ref{FollowGuide} panel b), more than half of all four classes stick to the instructor-only materials. The usage of instructor-only materials is dominated in PHY156, PHY116, PHY206, and PHY114 in order. ChatGPT-generated answers are provided mostly by PHY114. This can correlate with the learner's discipline. PHY114 students are from a health professional and this class is very different and they mostly feel it has no impact on their future carrier. It seems that they are not caring about their learning but just wanted to pass the class. On the other hand, ChatGPT is used as a learning tool mostly in PHY206. These are students from educational science and it seems that they care more about their learning rather than just passing the class. PHY116 and PHY156 students seem to following only instructor materials. These students are from the engineering or biological discipline and it seems they are more concerned about their learning because they know the importance of concepts and applications for the future. 
The results suggest a strong correlation between following instructor materials for learning and the student's discipline of studies. Also, it shows that ChatGPT can be used as an effective learning tool which means in the future we may be able to modify teaching methods to incorporate ChatGPT as a standard learning tool. 

\begin{figure}[!ht]
  \centerline{\includegraphics[width=0.50\textwidth]{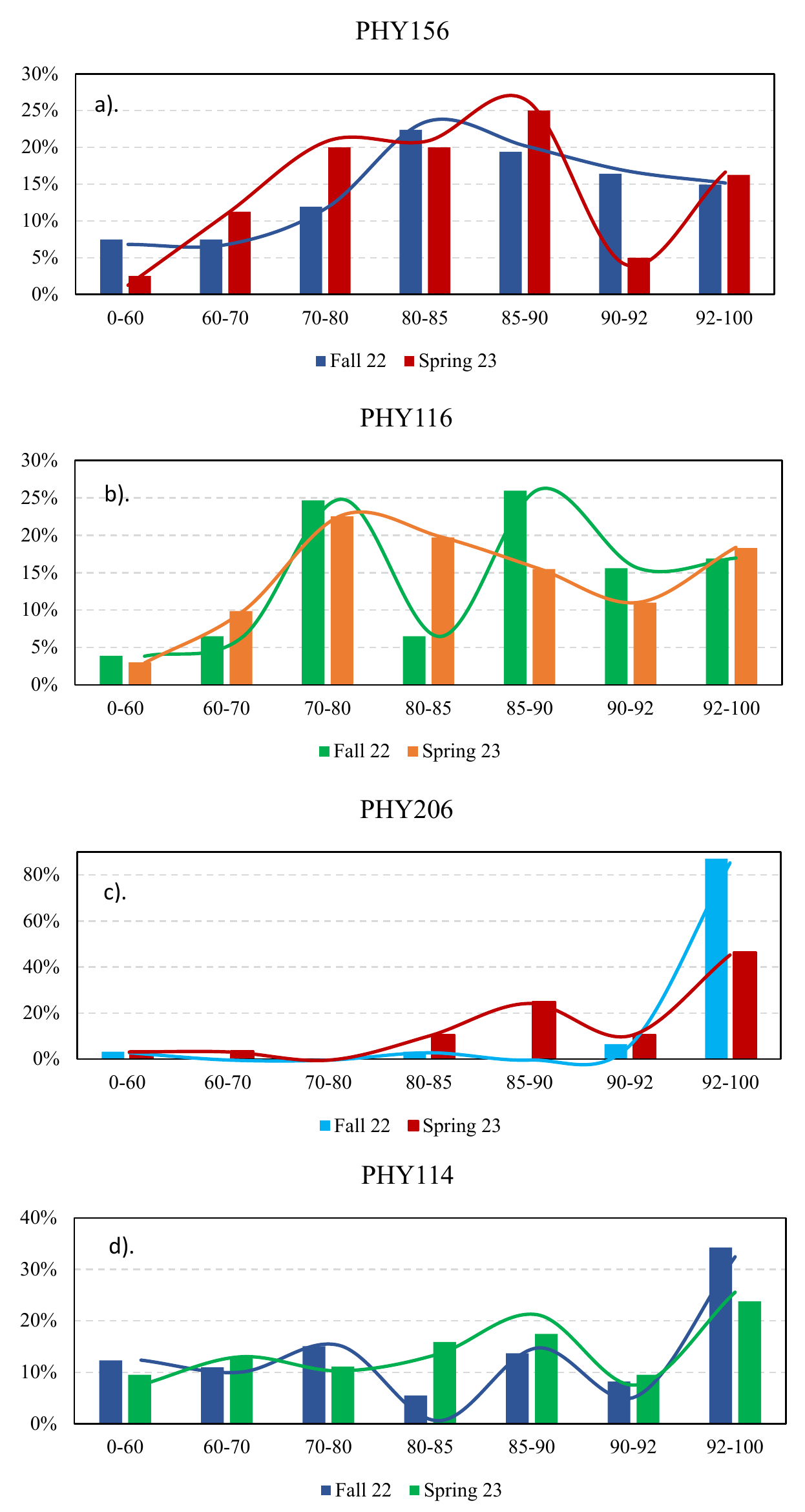}}
  \caption
    {(Color online) Lecture final grade analysis with mandatory and choice-based problem-solving sessions. Panel a) displays the PHY156 percentage of students who belong to noted grade distributions according to their performances by using blue for Fall 22 and red for Spring 23. Panel b, c, and d show the same results for PHY116, PHY114, and PHY206 respectively.}
\label{FinalGrade}
\end{figure}

We study the student outcomes by using mandatory synchronous problem-solving sessions (learners have to join online and do the work and submit), to choice-based problem-solving sessions (learners have a choice to join and complete the work or complete the work later by watching the recording). Additionally, we separate lecture notes assignments from homework assignments in Spring 23 to encourage students to use E-lecture more effectively. All the other structures stay the same. In the choice-based setting in Spring 23, less than 10\% of each class join the session synchronously, the others choose to watch the recording asynchronously. This suggests that students like to study on their own time than just assigned time schedules for synchronous sessions. 
We investigate the percentage of total final points for the lecture portion of all four classes and the results are shown in Figure~\ref{FinalGrade}. Since the laboratory part was conducted by different instructors and we did not include that in the comparison. Therefore, the results in Figure~\ref{FinalGrade} will not represent the final letter grade. All quizzes and final exams for PHY116 and PHY156 were proctored synchronously by the instructor in all semesters. PHY114 and PHY206 learners complete the timed quizzes and timed final exams asynchronously in all semesters. All the other settings stayed the same for all four classes and two semesters.

\begin{figure}[!ht]
  \centerline{\includegraphics[width=0.5\textwidth]{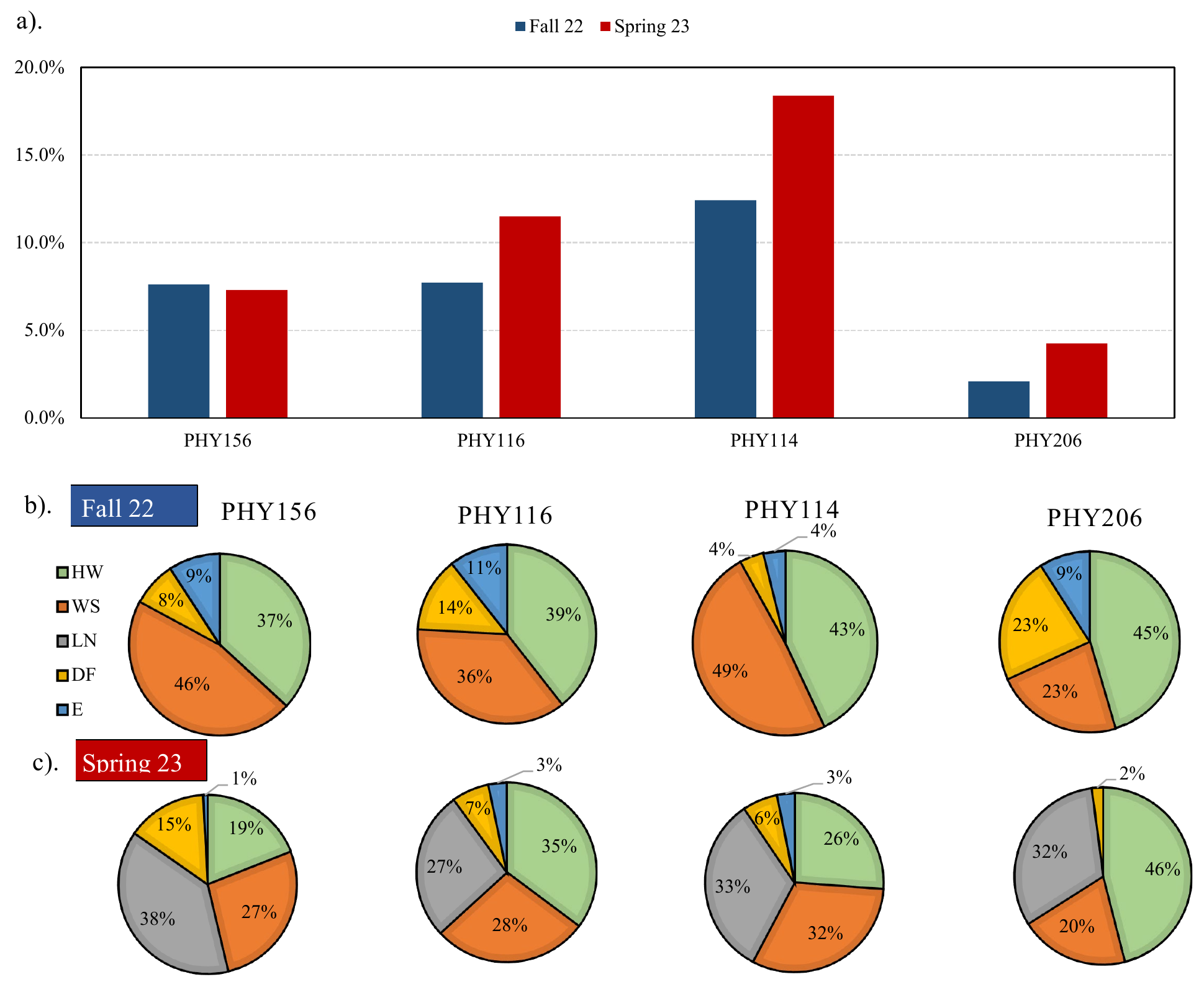}}
  \caption
    {(Color online) Percentage of missing assignments and breakdown of the missing assignments by the class. Panel a) represents the percentage of the missing number of assignments for each class by using blue for Fall 22 and red for Spring 23. Panels b) and c) show the breakdown of the missing assignments' percentage for each class for Fall 22 and Spring 23 respectively. Abbreviations are used as HW-homework, LN-lecture note, WS-worksheet, DF-discussion forum, E-Exam}
\label{Missing}
\end{figure}

In Spring 23, the number of students within the percentage range of 92-100 in PHY116 and 156 shows a slight improvement with choice-based problem-solving. The number of Students' percentages within the range of 0-60 also dropped in spring 23 in all four classes. In these two situations, we can conclude that choice-based problem-solving sessions may help the improvement. This further suggests that students like to work on their own time schedule and this is particularly important for most of the CSI students because most of them work part- or full-time while a full-time student. 
Since PHY206 and PHY114 learners completed all exams as timed asynchronously, the range of 92-100 students is dominated in both semesters, but the percentage of students in that range dropped in spring 23 by showing better grade distribution. This may be due to the fact that choice-based students forget to submit the required work for the sessions. The mandatory problem-solving session seems more effective in PHY206 and PHY114 to improve the students in the 92-100 range or it may be the coincidence of two semesters students' variation. The other ranges have deviated from one another, we are unable to discuss the effect of choice-based problem-solving sessions accurately there. The odds may be due to the student's effort of reaching a higher grade since the blackboard grade center allows learners to track their growth during the semester. Therefore, there is not much effect on the method of conducting problem-solving sessions on the student performances. And also overall grade distribution of all classes is good although the classes are online setting. That also indicates that there are not any considerable issues of academic dishonesty. Only the PHY206 class displays minor grade distribution, which may be due to the small class size and engagement with future educators. 

We study a number of missing assignments due to the effect of choice-based problem-solving and the increment in the number of submissions. Since we separate lecture notes assignments from homework assignments in Spring 23, there were 12 extra assignments added to the list. That means students in Fall 22 had 31 assignments and students in Spring 23 had 43 assignments altogether, but the workload was the same for both semesters. The percentage of missing assignments for each class is shown in Figure~\ref{Missing} for two semesters separately for all four classes. Panel a) represents the percentage of missing assignments for each class by denoting red for Spring 23 and blue for Fall 22. It shows that with the increasing amount of submissions, learners will miss about the same number of submissions for all classes except PHY156. This may be due to the fact that PHY156 students are more experience with the class structure because they already completed PHY116. 
The breakdown percentage of missing exams (denoted with E) is shown in panel b) for Fall 22 and panel c) for Spring 23. It can be observed that there are no missing exams in Spring 23, but there are a few who missed the exams in Fall 22. On the other hand, most missing works are weekly assignments such as homework, worksheets, and lecture notes.  

By posting anonymous surveys on Blackboard, students' envision was investigated in terms of the class structure and their learning. Answers are independent of the discipline of the class and more than 75\% of learners contributed voluntarily to the surveys. More than 90\% of learners prefer the class to be taught online. More than 95\% of learners mentioned the class structure and class hours effectively used. More than 80\% agree the labs and lectures are well-aligned. More than 95\% mentioned that they learn enough materials and were happy about the instructor's materials, zero-cost (OER) textbook, and feedback provided. More than 80\% mentioned the class required a reasonable amount of work that they can handle online. More than 90\% was happy about the class structure and the grade they earn.
By considering students' final feedback, performance, progress, and grade distributions we safely suggest that this restructured 3-component online course model is well suited for future teaching and learning of introductory-level courses. 

\section{Conclusion}
Introductory physics classes are restructured to create an effective online experience for undergraduate students. Restructured classes consist of three components: asynchronous E-lecturing, synchronous mandatory, and/or choice-based problem-solving, and in-person laboratories. All classes are well structured, and implemented with a learning management system (LMS). We measure students' learning outcomes and students' progress throughout the semester and extend the teaching methods over a couple of semesters. We use zero-cost (OER) materials, and LMS effectively to engage learners effectively throughout the semester.  By manual grading, providing individual feedback, and holding individual meetings, we allow students to engage and discuss their difficulties and struggles throughout the semester. We use ID-integrated breakdown problems to individualize assignments and the final conclusions on problem-solving. This provides personalized questions for every student and each of them solves different questions and writes personalized conclusions accordingly. Further, symbol-integrated, breakdown, and individualized questions help learners to understand the materials deeper and help instructors to investigate students learning. 

We investigate students' answers with ChatGPT-generated answers to check the effect of AI involvement in learning and academic dishonesty issues. Results suggest that the use of ChatGPT on assessments is very low (below 12\%) in three courses and close to 25\% in one particular course. Further analysis shows that the use of ChatGPT may correlate with students' majors and their attitudes toward learning Physics, other than the difficulty level of the courses. If students think that Physics is important for their future ( PHY156 and PHY116), the use of ChatGPT is minimum. Otherwise, it shows considerable use of ChatGPT (PHY114 and PHY206). Also, we speculate the percentage of ChatGPT usage is correlated with students' discipline, for example, PHY 114 and PHY 206 are following the same class materials, but education majors are willing to work without using ChatGPT. Our investigation doesn't show any ChatGPT correlation between gender identities, we need further investigation to conclude.

We implement various tasks and efforts to minimize academic dishonesty issues and help learners to reach the highest education and effective learning experience. This was implemented by imposing personalized class requirements to use class materials honestly and ID-integrated assessments which personalize the questions and conclusions for each student. It is also identified that the sustainability of assignments is straightforward. The results suggest that they are great tools to minimize academic dishonesty and also provide students with a personalized learning experience in online large-scale courses. We also found that choice-based problem-solving sessions will display the same student's performance as the mandatory attendance session, but allow students to complete their own work schedule based on individual preferences and circumstances.

Anonymously distributed survey results display exceptional feedback from learners about the redesigned class structure, learning materials, and overall satisfaction with learning. This class structure is well-suitable for current student populations in most universities mainly most students work part or full-time. This is further supported by the evidence of more than 75\% of students in all four classes work half or full-time while taking a full-time course load. On the other hand, the class model shows very high student performance and also an exceptionally high retention rate.  
Survey results also show that there is a detectable correlation between gender and student performance. More than half (close to 75\% in 2 courses) of the registered students are identified as female and student performance (particularly above 80\% in final grade) seems to correlate with gender, although it may need further investigation. However, there is one particular course (specially designed for education science majors) that shows exceptionally high student performance and more than 90\% of students are female. 

The study suggests that the restructured online class model with student college ID-integrated assessments and personalizing the class and class materials is an efficient model to deliver and evaluate online courses in any science discipline. 

\section*{Acknowledgments}
We acknowledge the College of Staten Island and Prof. Charles Liu (chair of the Department of Physics and Astronomy) for allowing us to remodel the classes and collect all necessary data for this study. This research received no financial support. 

\section*{Disclosure statement}
No potential conflict of interest was reported by the authors.

\section*{Ethical Statement}
This research was done according to ethical principles outlined by the Institutional Review Board (IRB) at the College of Staten Island. The project didn't require IRB review because the manuscript was determined to be an educational quality improvement initiative, not research with human subjects. This article does not contain any personal data, and it does not identify any individual. All students have given explicit consent to participate in this research project.
\bibliographystyle{IEEEtran}
\bibliography{main.bib}
\end{document}